
\magnification=\magstep1
\def\etal{\it et al \ \rm}
\def\kms{km s$^{-1}$}

\def\gsim{ \lower .75ex \hbox{$\sim$} \llap{\raise .27ex \hbox{$>$}} }
\def\lsim{ \lower .75ex \hbox{$\sim$} \llap{\raise .27ex \hbox{$<$}} }
\def\pp{\noindent\parshape 2 0truecm 16.0truecm 1.0truecm 15truecm}

\def\spose#1{\hbox to 0pt{#1\hss}}
\def\simlt{\mathrel{\spose{\lower 3pt\hbox{$\mathchar"218$}}
     \raise 2.0pt\hbox{$\mathchar"13C$}}}
\def\simgt{\mathrel{\spose{\lower 3pt\hbox{$\mathchar"218$}}
'     \raise 2.0pt\hbox{$\mathchar"13E$}}}
\overfullrule=0pt
\font\twelverm=cmr10 at 12truept  
\twelverm    
\baselineskip=12.045truept 
\parindent=2.5em     
\parskip=0pt

\hrule height0pt
\baselineskip 21pt 	
\baselineskip 12pt
\parskip=3pt
\parindent=36pt

\hsize=6.5truein
\vsize=8.truein

\font\titlefont=cmss17

\vskip 0.5truein

\centerline{\titlefont
Constraints on the global mass to light ratios and extent of}
\vskip 0.1truein
\centerline{\titlefont
dark matter halos in globular clusters and dwarf spheroidals}

\vskip 0.8truein
\centerline{\bf Ben Moore}
\vskip 0.2truein

\parindent=0pt
\centerline{\it Department of Astronomy,
University of California, Berkeley, CA 94720, USA.}

\parindent=36pt

\vskip 1.0truein


\centerline {\bf Abstract}
\vskip 8pt
\parindent=36pt
\vskip 0.3truecm

\noindent

The detection of stars in the process of being tidally removed from
globular clusters and dwarf spheroidals in the Galaxy's halo provides
a strong constraint on their mass to light ratios and on the extent of
their possible dark matter halos. If a significant dark matter
component existed either within or beyond the observed stellar
distribution, then stars would not be removed. We use numerical
simulations to study mass loss from two component star clusters
orbiting within a deeper potential. We find a global upper limit on
the mass to light ratios of globular clusters, M/L$\lsim 2.5$, and rule out
the possibility that they have extended halos of low luminosity
material.  Similarly, the tidal tails of dwarf spheroidals indicates
that their dark matter halos must be truncated at $\sim 400$ pc
therefore they have total mass to light ratios $\lsim 100$.

\vskip 20pt
\vskip 20pt
\noindent{\bf Subject headings.} \ {\it Dark Matter, Galaxy: Formation.
Globular clusters: general. Stars: low mass, brown dwarfs. Galaxies:
halos, interactions. }

\vfil\eject

\vskip 20pt
\centerline {\bf \S 1. Introduction}
\vskip 5pt
\parindent=36pt

Dark matter plays an important role in the dynamics of a wide range of
galactic systems with masses ranging from $10^7M_\odot$ to
$10^{15}M_\odot$. Comprehension of the nature and distribution of dark
matter is crucial to understanding the evolution of the Universe.
Dwarf spheroidal galaxies are only a few hundred parsecs across and
are the smallest systems in which dark matter has been detected
(Aaronson 1983). These galaxies have the highest known dark matter
densities $\sim 1M_\odot$ pc$^3$ and the motions of their stars are
completely dominated by dark matter at all radii. Their luminosities
are similar to globular clusters, however their sizes are quite
different. Globular clusters have core radii $r_c\sim 2$ pc, two
orders of magnitude smaller than the dwarf spheroidals. Although the
magnitude of the observed stellar velocity dispersions are similar,
the dynamical analyses indicate mass to light ratios M/L$\sim
0.5\--2.5$ in globular clusters (Pryor \etal 1989) and M/L$\sim
30-100$ in dwarf spheroidals (Lake 1990).

The stellar velocities measured for several globular
clusters show a steady decline in dispersion with radii, although
they never fall to zero.
The possibility remains that the luminous
component is embedded within a larger sub-luminous dark halo that
could extend well beyond the observed ``tidal radii'' of the cluster
({\it c.f.} Heggie \& Hut 1995, and references within).
Extrapolation of the observed mass function of stars in globular
clusters have lead to claims that these systems may have large
populations of low mass faint stars such as brown dwarfs (Fahlman
\etal 1989, Richer \etal 1991, Taillet \etal 1995).
Mass segregation would help populate an extended halo of dark matter
around clusters and thus the measured M/L ratios are only local values
and much higher global values are possible.

Alternative models for dark matter dominated clusters have been
proposed by other authors (Peebles 1984, Rosenblatt \etal 1988).
Peebles suggests that two natural scales exist in hierarchical structure
formation models in which the mass density is dominated by cold dark
matter.  In this scenario globular clusters form naturally within the
extended dark halos of weakly interacting particles. The baryons can
dissipate energy which may result in significant segregation between
the light and the mass, leading to global values of M/L as
large as 100.

Recent observations by Grillmair \etal (1995) and Irwin \&
Hatzidimitriou (1993) have revealed the presence of symmetric tidal
tails of stars being gravitationally removed from globular clusters
and dwarf spheroidals. In this {\it letter} we explore the
consequences of these observations and how they can be used to
constrain a possible component of dark matter within or around these
objects. Stellar systems containing dark matter will have larger tidal
radii and higher escape velocities, hence the rate of mass loss via
evaporation and tidal heating will be lower. We use N-body simulations
of star clusters orbiting within a galactic potential to determine the
maximum amount of dark matter that could be present in order to be
consistent with the observed rate of mass loss.

\vskip 20pt
\centerline {\bf \S 2. Tides and Tails}
\vskip 5pt
\parindent=36pt

Globular clusters and dwarf spheroidals orbit within the massive dark
halo of the Galaxy. These systems are subject to a tidal
compression along an axis pointing towards the center of the deeper
potential.  At a certain distance from the system, a contour of
equipotential exists, such that stars will become bound to the Milky
Way if they pass beyond this critical surface.  Stars will tend to
leave the satellites via the extremities of the tidally extended
system. Those stars that leave the satellite furthest from the Galaxy
will begin to lag behind since their orbital timescale is
lower. Likewise, the stars that leave the satellite closest to the
Galaxy will form a leading tail due to their higher energies. The
stars leave the cluster at close to zero relative velocity, therefore
they will follow the same orbital path as the cluster within quite
narrow tails and in a direction perpendicular to the tidally extended
axis (e.g. Figure 1).

Globular clusters that orbit beyond the galactic disk suffer mass loss
primarily via evaporation as stars reach the escape velocity through
encounters. The evaporative mass loss from a truncated star cluster can
be calculated (e.g. Johnstone 1993 and references within) and is
typically between 1\---10\% per half mass relaxation time:
$$
\eqalignno{
T_{_{1/2}} &= {
(2/3)^{1/2}(<v^2>)^{3/2}
\over
(M/2V_h)<m>4\pi G^2 {\rm ln}\Lambda
} \ ,
&(1)\cr}
$$
where $<v^2>$ is the mean-square speed of the stars, $<m>$ is the
mean mass per star, $M$ is the total mass, $V_h$ is the half mass
volume and $\Lambda$ is the ratio of maximum to minimum encounter
distances to be considered (Spitzer 1987).

Over its 15 billion year lifetime, a typical galactic cluster may have
lost up to half of its mass via evaporation alone. This stellar debris
is detectable using deep imaging observations since the tidal
tails are narrow. Grillmair \etal (1995) use photographic plates
to perform star counts beyond the observed tidal radii of a sample of
12 globular clusters. Symmetric tidal tails of escaping stars were
detected and as we shall show using N-body simulations, the observed
escape rates are roughly consistent with the analytic calculations.

The observed ``edges'' of the globular clusters are close to the
equipotential surface imposed by the Milky Way when mass to light
ratios of order unity are adopted.  Various dynamical processes will
enable stars to move past this surface and become lost to the
galaxy. The rate of mass loss depends upon the fraction of stars that
have enough energy to pass beyond this tidal radius. For a cluster of
fixed physical size, larger mass to light ratios will increase the
escape velocity and slow the evaporation rate.  As stars diffuse to
larger radii they pass through the higher density core less frequently
and suffer fewer encounters, thus their ability to reach the escape
velocity rapidly decreases.

If the dark matter is distributed on a larger scale than the luminous
component, e.g. when the stars are sitting deep within the core of a
larger dark matter halo, then the ability of the visible stars to
diffuse to large radii would be reduced. In this case the mass
to light ratio within the visible extent of the cluster may not be
greatly increased, but the total mass could be many times greater than
the mass inferred from the stellar dynamics within the optical
radius. Consequently, even modest amounts of dark matter will be very
effective at containing the visible stars and halting the production
of tidal tails.

One caveat is that highly radial orbits might strip a pre-existing
dark matter halo, however, many globular clusters have proper motions
determined (Cudworth \etal 1993).  For example, M2 (NGC 7089) has very
prominent tidal tails, but the measured space velocity would take the
cluster on an orbit between 8 kpc and 15 kpc within an isothermal
potential.  The tidally imposed radius, $R_t$, of a system orbiting at
a distance $R_G$ within a deeper potential is the point at which their
mean densities are equal. For point masses
$R_t=R_{G}(m_{globular}/3M_{Galaxy})^{1/3}$, we find tidal radii for
M2 of 81 pc for M/L=1 and 117 pc for M/L=3.  Here we have taken the
luminosity of M2 to be $L_B=2.5\times10^5L_\odot$ and the mass of the
Milky Way as $10^{11}M_\odot (R_G/10{\rm kpc})$.  These values should
be compared with its King model tidal radius of 60 pc. If the observed
stars of M2 were moving within the core radius of a dark halo with an
isothermal density profile, then the halo would be tidally limited at
$R_t\approx R_{G}\sigma_{Galaxy}/\sigma_{halo}$. For
$\sigma_{halo}=10$\kms\ and $\sigma_{_{Galaxy}}=150 $\kms\ we find
that the globular's dark halo could extend to 500 pc and its global
M/L$\sim 50$.  In order to strip such a halo from M2 would require a
pericentric distance of $\sim 1$ kpc.

The detection of stars being tidally removed from these systems
provides strong evidence that their physical extent is close to that
indicated by the stellar distribution. In the next section we shall use
N-body simulations to support this discussion and to provide more
quantitative numbers on the rate of mass loss from stellar systems
that contain dark matter.

\vskip 20pt
\centerline {\bf \S 3. Constraining the global distribution of dark matter}
\vskip 5pt
\parindent=36pt

We construct equilibrium globular clusters by sampling a King profile
with core radius 3 pc and tidal radius 60 pc.  Our standard cluster
contains no dark matter and has a total mass of $2\times 10^5M_\odot$
represented by 10,000 particles.  Two additional models were
constructed such that the stellar component was embedded within a dark
matter halo.  The dark matter was distributed using a lowered
isothermal sphere with core radius 30 pc and an exponential cutoff at
75 and 100 pc.  The mass of the dark particles was set to 1/4 of the
star particles.  These models had M/L values of 1.5 within the optical
radius but the global values were 2 and 3 respectively.

The clusters were placed in orbit within a fixed isothermal potential
of circular velocity 220 \kms, such that the apocentric and
pericentric distances were 15 kpc and 8 kpc respectively.  We evolved
the particle distributions using the TREECODE (Barnes \& Hut 1986)
with tolerance 0.7 and enough time-steps such that each particle
travels 3 steps across the softening length of 1.5 pc at a velocity
equal to the central dispersion $\sim 10$\kms.

We ran our standard cluster with no dark matter for 10 Gyrs and we
found that the mass loss is roughly constant over this period.  For
this model, $T_{_{1/2}}\sim 2.5\times10^8$ years and for a single mass
model with these structure parameters the rate of mass loss is about
1\% per half mass relaxation time (Johnstone 1993).  The amount of
mass unbound from the cluster at the end of this calculation was 30\%,
in reasonable agreement with the analytic estimate. Note that this
example resembles the globular cluster M2, but that our results are
easily generalised to any star cluster orbiting within a deeper
potential.

Simulating a cluster for a Hubble time, even at our limited resolution
requires more than $10^5$ timesteps. However, since the rate of mass
loss is constant we can compare our simulations with the observations
of Grillmair \etal after only a few Gyrs.  Figure 1 shows a snapshot
of a model with dark matter after 4 Gyrs.  The tidal tails of dark
matter are clearly visible since roughly 15\% of the dark matter has
reached the cluster escape velocity via relaxation of our artificially
massive dark matter particles.  Low mass stars or WIMPS will produce
much smaller tidal tails since the relaxation timescale is
proportional to ${<m>}^{-1/2}$.

The imaging data of Grillmair \etal reveal of order a hundred stars
above the background that are escaping from the cluster. This
represents less than 1\% of the total number of visible stars that are
presently bound to the cluster.  This is roughly the expected number
of stars in the surveyed area for a typical cluster with no dark
matter calculated using the evaporation rates from Johnston (1993) for
a multi-component mass model.

In order to compare with the observations we smooth our data on a
similar scale as Grillmair \etal and plot iso-density contours of the
stars.  Figure 2 shows our results for the 3 models.  Clearly, even
modest amounts of dark matter can slow down or halt the stellar mass
loss.  The tidal tails from our standard model are quite similar to
the observations, however, as the M/L ratios are increased the tails
become much less prominent.  Tidal extensions are still visible in our
model which has a global M/L=2, although the stellar mass loss has
been reduced by about a factor of 5 over the standard model with
M/L=1.  When we increased the mass to light ratio to 3, then the
stellar mass loss is reduced to zero. From these simulations we
conclude that the total mass to light ratios of globular clusters
cannot be any higher than $\sim 2.5$, otherwise escaping stars would
not have been detected.  (Similar results were obtained when we
distributed the dark matter in an identical way to the stars.)

\noindent{\bf 3.1 Numerical artifacts}

How are these results effected by numerical resolution?  We are using
10,000 mass points of effective size 1.5 pc to represent a star
cluster that has an order of magnitude more stars of much smaller
physical size.  Evaporation is a diffusion process that is dominated
by distant encounters rather than close, strong scattering events.
Therefore the fact that our stars are physically softened does not
reduce the evaporation rate significantly. A more important artifact
is a higher than expected mass loss rate since our star particle mass
is effectively 20$M_\odot$. This can only serve to underestimate our
constraint on the global M/L values.  Our treatment of the dark matter
also enhances the evaporation rate of star particles since the dark
particles are also artificially massive.  As discussed earlier, in
real clusters the dark matter is expected to be much lighter than the
visible stars, thus we may be underestimating the effects of mass
segregation. We do find that the half mass radius of the stars slowly
decreases, although the rate of this effect is significantly smaller
than analytic calculations yield (Spitzer 1987). More detailed
simulations or analytic calculations could yield stronger constraints
on the global M/L values.

\vskip 20pt
\centerline {\bf \S 4. Discussion and conclusions}
\vskip 5pt
\parindent=36pt

Symmetric tidal tails of stars being gravitationally stripped from
globular clusters indicates that their mass distributions do not
extend beyond the optical radius. This rules out the possibility of
large dark matter halos surrounding globular clusters and hence
excludes the possibility that these objects formed as density peaks
within a cold dark matter type scenario (Peebles 1983, Rosenblatt
1988).  We can also use these results to rule out the presence of
large quantities of low mass stars either within or beyond the optical
radii (Fahlman \etal 1989, Richer \etal 1991, Taillet \etal 1995). A
comparison of stellar isophotes beyond the tidal radii with the data
of Grillmair \etal (1995) provides an upper limit on the total mass to
light ratios of globular clusters of M/L $\lsim 3$.

Tidal tails have also been observed to extend from the Ursa-Minor
dwarf spheroidal (Irwin \& Hatzidimitriou 1993).  Data from these
authors shows a large number of stars beyond the classical tidal
radius of 400 pc.  Our results can be readily applied to the dwarf
spheroidal galaxies and suggests that the dark matter halo of
Ursa-Minor must be truncated at the optical radius $\sim 400$ pc and hence
the mass distribution traces the light distribution.

If the dwarf spheroidal galaxies formed within cold dark matter halos
then the fact that they are truncated at $\sim 400$ pc has interesting
consequences for cosmology. Namely, for the first time
we have obtained an {\it upper} limit on the mass to light ratio of a
dark matter dominated system.  The radial velocity dispersion of stars
within dwarf spheroidals appears to be fairly constant with radius If
we parameterise the halo by an isothermal sphere with 1-d velocity
dispersion 10 \kms, then the virial radius should nominally extend to
$15$ kpc for a closed Universe.  Tidal truncation at the satellites
present galacto-centric distance would limit the radial extent of its
halo to $\sim 5$ kpc. In order to strip the halo to within 400 pc
would require a pericentric distance of $\sim 6$ kpc.  Proper motion
studies of this galaxy would be extremely interesting and could be
used to distinguish between halo formation within a low density
Universe, or tidal stripping on an elongated orbit.

\vskip 5pt
\centerline {\bf Acknowledgments}
\vskip 5pt

I would like to acknowledge useful discussions with Fabio Governato
and Doug Johnstone. The numerical simulations were performed and
analysed using the resources of the HPCC group at Seattle, funded
by NASA through the LTSA and
HPCC/ESS programs.

\vskip 5pt
\centerline {\bf References}
\vskip 5pt

\pp Aaronson M. 1983, {\it Ap.J.Lett.}, {\bf 266}, L11.

\pp Barnes J. \& Hut P. 1986, {\it Nature}, {\bf 324}, 446.

\pp Cudworth K.M. \& Hanson R.B. 1993, {\it A.J.}, {\bf 105}, 168.

\pp Fahlman G.G, Richer H.B., Searle L. \& Thompson I.B. 1989,
{\it Ap.J.Lett.}, {\bf 343}, L49.

\pp Grillmair C.J., Freeman K.C., Irwin M. \& Quinn P.J. 1995,
{\it A.J.}, {\bf 109}, 2553.  \etal

\pp Heggie D.C. \& Hut P. 1995, {\it I.A.U. Symposium 174},
Dynamical Evolution of Star Clusters - Confrontation of Theory
and Observations.

\pp Irwin M.J. \& Hatzidimitriou D. 1993, {\it ASP Conference series},
{\bf 48}, 322.

\pp Johnstone D. 1993, {\it Ap.J.}, {\bf 105}, 155.

\pp Jones B.F., Klemola A.R. \& Lin D.N.C. 1994, {\it A.J.}, {\bf 107}, 1333.

\pp Lake G. 1990, {\it M.N.R.A.S.}, {\bf 244}, 701.

\pp Oh K.S., Lin D.N.C. \& Aarseth S.J. 1995, {\it Ap.J.}, {\bf 442}, 142.

\pp Peebles P.J.E. 1984, {\it Ap.J.}, {\bf 277}, 470.

\pp Pryor C., McClure R.D., Fletcher J.M. \& Hesser J.E. 1989,
{\it Ap.J.}, {\bf 98}, 596.

\pp Richer H.B., Fahlman G.G, Buonanno R., Pecci F.F \etal 1991,
{\it Ap.J.}, {\bf 381}, 147.

\pp Rosenblatt E.I., Faber S.M. \& Blumenthal G.R. 1988, {\it Ap.J.},
{\bf 330}, 191.

\pp Spitzer L. 1987, Dynamical Evolution of Globular Clusters,
(Princeton Univ. Press.), Princeton series in astrophysics.

\pp Taillet R., Salati P. \& Longaretti P.-Y. 1995, {\it Nuc.Phys.B.Supp.},
{\bf 43}, 169.

\vskip 5pt
\noindent{\bf Figure captions}
\vskip 5pt
\parindent=36pt

\noindent{\bf Figure 1.\ }
The particle distribution from one of our simulations after 4 Gyrs of
evolution. The viewpoint is looking down upon the orbital plane and
the curve depicts the cluster's orbit over 0.5 Gyrs.  In this
simulation the model globular cluster has a dark halo that extends
beyond the stellar component.  The presence of the dark halo has
reduced the stellar mass loss to practically zero. The tidal tails
contain 10\% of the dark matter particles which are lost primarily due
to relaxation because of their artifically large masses.  The inset
plot shows the distribution of stars which shows no evidence of mass
loss into tidal tails.

\noindent{\bf Figure 2.\ }
Contour plots of the stellar distribution in the immediate vicinity
of the globular cluster which should be compared with Grillmair \etal,
in particular their Figure 13. The axis scale is in parsecs and the
stellar density has been smoothed in projection using a Gaussian of
width 5 pc.  The arrows indicate the direction to the center of the
galactic potential.
(a) The standard model cluster without dark matter. (b) and (c) are
for the model cluster with a dark matter halo such that the total mass
to light ratios are 2 and 3 respectively.  The similarity between the
observations and the mass loss rate from our standard cluster in (a)
provides conformation that our understanding of stellar dynamics and
the evaporation process in star clusters is close to being correct.

\bye